\documentstyle[manuscript,prl,aps]{revtex}
\begin{document}
\def\hb{\bar{H}}
\def\bs{\bar{psi}}
\title{\bf DEGENERACY STRUCTURE OF THE CALOGERO-SUTHERLAND MODEL: AN 
  ALGEBRAIC APPROACH}
\author{N. Gurappa, Prasanta K. Panigrahi$^\dagger$, V. Srinivasan$^{*}$}
\address{School of Physics, University of Hyderabad,
Hyderabad - 500 046, {\bf India}}
\maketitle
\begin{abstract}
The degeneracy structure of the eigenspace of the N-particle
Calogero-Sutherland model is studied from an algebraic point of view.
Suitable operators satisfying SU(2) algebras and acting on the
degenerate eigenspace are explicitly constructed for the two particle
case and then appropriately generalized to the $N$-particle model.
The raising and lowering operators of these algebras connect the states,
in a subset of the degenerate eigenspace, with each other. 
\end{abstract}
\vfill
$\dagger$ E-mail: panisp@uohyd.ernet.in\\
{*} E-mail: vssp@uohyd.ernet.in
\newpage
The Calogero-Sutherland model (CSM),$^1$ describing $N$, interacting,
identical particles in one-dimension, have currently attracted wide interest,
both in physics as well as mathematics literature. These exactly solvable
quantum mechanical models have relevance to two-dimensional gravity,$^2$
quantum Hall effect,$^3$ fractional statistics$^4$ and a host of other
problems of physical interest.$^5$ For the model describing particles on a
circle interacting via the inverse sine square potential, the basis set of
the orthonormal eigenfunctions are the well-known Jack polynomials.$^6$ These
functions have been constructed recently, through an operator method,$^7$ by
making use of the $S_N$-{\it extended} Heisenberg algebra.$^8$ However, for
particles on a line, interacting through pair-wise inverse square potential
in the presence of harmonic confinement, the complete set of orthonormal
states is yet to be constructed explicitly. In light of the degeneracy of
this model, it is of considerable interest to provide operators that commute
with the Hamiltonian and connect the members of the degenerate eigenspace
(DE). One would also like to unravel the underlying algebraic structure of
these operators.

For the two particle case, this problem has been recently analyzed and the
symmetry, associated with the degeneracy, has been shown to originate from a
polynomial extension of the SU(2) algebra.$^9$ In this paper, we first
re-analyze the two particle CSM and show that the degeneracy can be described
by a SU(2) algebra. The above method is then extended to the $N$-particle
case, where we point out the existence of several copies of $SU(2)$ algebras.
The raising and lowering operators of these algebras connect various
sub-sectors of the DE. The advantage of this approach lies in the fact that,
for more than two particles, the polynomial $SU(2)$ algebra becomes fairly
complicated and unlike the $SU(2)$ case, one needs to find the unitary
representations of this algebra individually for each case. However, we have
been unable to find the complete set of operators, that will connect all the
degenerate states with each other.

The CSM Hamiltonian is given by  $( \hbar = \omega=m=1)$
\begin{equation}
H  =  -{1\over2} \sum_{i=1}^N \partial_i^2 +  {1\over2}  \sum_{i=1}^N  x_i^2
+ {g^2\over2} \sum_{{i,j} \atop {i \ne j}}^N {1\over(x_i-x_j)^2} \,\,\,\,.
\end{equation}

We work in a sector of the configuration space corresponding to a definite
ordering of the particle coordinates: $x_1 \le x_2 \le \cdots \le x_N$.
After performing a similarity transformation (ST), $H$ takes the following
form,
\begin{eqnarray}
{}H^\prime &=&  Z^{-1} H Z \nonumber\\ &=& -{1\over2} \sum_i  \partial_i^2  +
\frac{1}{2} \sum_i x_i^2 - \lambda \sum_{{i,j}\atop{i\ne j}}
{1\over(x_i-x_j)} \partial_i  \qquad.
\end{eqnarray}
where $Z =  \prod_{i<j}^N [|(x_i-x_j)|^\lambda  (x_i-x_j)^\delta]$, and $g^2
= (\lambda + \delta) (\lambda +\delta - 1) $. Here $\delta = 0$ or $1$
represents the choice of the quantization of the $N$-particle system as
bosons or fermions respectively; we choose $\delta = 0$ for convenience.

On symmetrized eigenstates, the above Hamiltonian $H^\prime$ is equivalent to
another one, $\bar{H}$, which can be factorized by the raising and lowering
operators of the $S_N$-extended Heisenberg algebra (SEH).$^8$ SEH is given by
$[a_i, a_j]  =  [a^\dagger_i, a^\dagger_j] = 0$ and $[a_i, a_j^\dagger]  =
\delta_{ij}\{1+\lambda \sum_l  K_{il}\} - \lambda K_{ij} $;
where, the transposition operator $K_{ij}$ satisfies
\begin{eqnarray}
{}K_{ij} & = &  K_{ji}\,\,\,;\qquad  (K_{ij})^2 = 1 \,\,\,,\nonumber\\
{}K_{ij}a_j &  = & a_iK_{ij}\,\,\,;\qquad K_{ij}a^\dagger_j =
a^\dagger_iK_{ij} \qquad  \mbox{(no summation over repeated indices)},
\nonumber\\
{}K_{ij}K_{jl}  & =&  K_{jl}K_{il}  =    K_{il}K_{ij}  \,\,\,,
\qquad  \mbox{for} \,\,\,i\ne j,\,\, i\ne l,\,\, j\ne l \,\,\,,\nonumber\\
{}K_{ij}K_{mn} & =&  K_{mn} K_{ij} \,\,\,,\,\,\ \mbox{for} \,\,\,  i,j,m,n
\,\,\, \mbox{all different}.
\end{eqnarray}
Explicitly, $$a_i (a_i^\dagger)  =  {1\over\sqrt{2}} (x_i + (-)D_i) \,\,\,,$$
where, $$D_i  = \partial_i + \lambda  \sum_{j\atop{j \ne i}}^N
{1\over(x_i-x_j)} (1-K_{ij}) $$ is known as the Dunkl derivative.$^{10}$ Now
\begin{eqnarray}
\hb  =  {1\over2} \sum_i^N \{a_i, a_i^\dagger\}
& = &\left(-{1\over2}\sum_i^N \partial_i^2 + {1\over2} \sum_i^N x_i^2 -
\lambda \sum_{{i,j}\atop{i\ne j}}^N  {1\over(x_i-x_j)}
\partial_i \right) \nonumber\\
& + &  \frac {\lambda}{2} \sum_{{i,j}\atop{i\ne j}}^N {1\over(x_i-x_j)^2}
(1-K_{ij})\nonumber\\ & - & \frac {\lambda^2}{2}\sum_{{i,j,l}\atop{i\ne {j ,
l}}}^N {1\over(x_i-x_j)} (1-K_{ij}) {1\over(x_i-x_l)} (1-K_{il})
\end{eqnarray}
and $[\hb, a_i (a^\dagger)]  =  - a_i (a^\dagger_i)$.  It is quite obvious
that on symmetric states, the last two terms in Eq. (4) vanish identically
and hence $\hb = H^\prime$.

Recently, it has been shown in Ref. 9 that the degeneracy of the two particle
CSM can be explained by a polynomial $SU(2)$ algebra.  This was achieved by
separating the relative and center of mass, creation and annihilation
operators $A_1, A_1^\dagger$ and $A_2, A_2^\dagger$ respectively. The
Hamiltonian in Eq. (1) can be written as $$H = {1\over2} (\{A_1,A_1^\dagger\}
+  \{A_2,A_2^\dagger\}) = H_1 + H_2\,\,\,,$$ where $[A_1,A_1^\dagger] = 1 + 2
\lambda \sigma$, $[A_2,A_2^\dagger] = 1 $ and all other commutators vanish.
Here $\sigma \equiv K_{12}$ anticommutes with $A_1, A_1^\dagger$ and commutes
with $A_2$ and $A_2^\dagger$. Notice that, $\frac{1}{2}A_1^2,
\frac{1}{2}A_1^{\dagger 2}$ and $\frac{1}{2} H_1$ generate a $SU(1,1)$
algebra; this is a part of the spectrum generating algebra of the two
particle CSM. An un-normalized, generic, symmetric eigenstates can be written
as
\begin{equation}
{}|n_1,n_2> = A_1^{\dagger 2{n_1}} A_2^{\dagger {n_2}} |0>\qquad,
\end{equation}
where $|0>$ is the ground state satisfying $A_1 |0> = A_2 |0> = 0$.  The
generators of the polynomial $SU(2)$ algebra, which maps symmetric states
into symmetric states are,
\begin{eqnarray}
{}J_o &=& {1\over4} (A_1^\dagger A_1 - A_2^\dagger A_2) \nonumber\\ {}J_+ &=
& \frac{\sqrt\alpha}{8} (A_1^\dagger)^2 A_2^2 \nonumber\\ {}J_-
&=&\frac{\sqrt \alpha}{8}(A_2^\dagger)^2 (A_1)^2\,\,\,.
\end{eqnarray}
They commute with the Hamiltonian and satisfy
\begin{eqnarray}
{}[J_o , J_{\pm}] &=& {\pm} J_{\pm}\nonumber\\ {}[J_+ , J_-] &=& 2 J_o -
\alpha J_o^3 - \beta J_o^2 + \gamma
\end{eqnarray}
Here, $\alpha = 32 {[N^2 + 2 (\lambda +1) N + 2 (\lambda^2 - 1)]}^{-1}, N =
A_1^\dagger A_1 + A_2^ \dagger A_2, \gamma = {1\over {64} } \lambda \alpha N
(N + 4)$ and $\beta = {3\over4} \lambda \alpha $.
 
For the above case, we first point out the method of construction of a
regular $SU(2)$ algebra underlying the degeneracy. This is done by making use
of the identities valid on the Fock space.$^{11}$ This procedure is then
generalized to the $N$ particle case.

In the following, we give the steps necessary for the construction of the
$SU(2)$ generators. For the operators $F_l = A_l^2$, $l = 1,2$, one can
construct the canonical conjugates (CC) $G_l^\dagger$: $[F_l , G_l^\dagger]
= 1$ (no summation over $l$) in several sectors. Explicitly, $G_l^\dagger$
can be written, in the vacuum sector as,$^{11}$
\begin{equation}
G_l^\dagger = {1\over2} F_l^\dagger \frac {1}{F_l F_l^\dagger}(A_l^\dagger
A_l + 2)
\end{equation}
The $SU(2)$ generators can then be defined as
\begin{equation}
{}J^+ = G_{1}^\dagger F_2 \,\,\,, {}J^- = G_{2}^\dagger
F_1 \,\,\,,\nonumber\\
\end{equation}
\begin{equation}
{}J^o = \frac{1}{2} (G_{1}^\dagger F_1 - G_{2}^\dagger
F_2) = \frac{1}{2} (A_1^\dagger A_1 - A_2^\dagger A_2)\qquad, 
\nonumber
\end{equation}
such that
\begin{eqnarray}
{}[J^0 , J^\pm] &=& \pm J^\pm \,\,\,,\nonumber\\ {}[J^+ , J^-] &=& 2 J^o
\qquad.
\end{eqnarray}
It is straightforward to check that $J^+, J^-$ and $J^0$ commute with $H$ and
act in the degenerate space.  The above procedure can be immediately
generalized to the $N$-particle case.  By choosing the center-of-mass,
relative coordinates and their differential operators, respectively as, $$ X
= \frac{1}{N} \sum_i x_i\,\,\,;y_i = x_i - X \,\,\,$$ and $$\partial_X =
\frac{1}{N} \sum_i
\partial_{x_i}; \partial_{y_i} = \partial_{x_i} - \partial_X$$
$H^{\prime}$ in Eq. (2) can be written in the form
\begin{eqnarray}
H^\prime  &=& \left( - \frac{N}{2} \partial_X^2 + \frac{N}{2} X^2\right) +
\left( -{1\over2}\sum_i^N  \partial_{y_i}^2 + \frac{1}{2} \sum_i y_i^2  -
\lambda \sum_{{i,j}\atop{i\ne j}}^N  {1\over(y_i - y_j)}  \partial_{y_i}
\right)\nonumber\\ 
 & = & H_X + H_{\{y_i\}}\qquad.
\end{eqnarray}
We define $A = \sqrt{\frac{N}{2}}(X+\partial_X)$ and $A^\dagger =
\sqrt{\frac{N}{2}}(X-\partial_X)$ 
such that
\begin{equation}
[A,A^\dagger] = 1\,\,,\,\,{} H_X = \frac{1}{2}\{A,A^\dagger\}\qquad,
\end{equation}
and $$[H_X,A(A^\dagger)] = -A(A^\dagger)\,\,\,.$$

$H_{\{y_i\}}$ can be factorized by making use of the SEH by replacing $x_i$'s
by $y_i$'s in Eq. (4). The modified commutator is
\begin{equation}
[a_i, a_j^\dagger]  =   \delta_{ij}\{1+\lambda \sum_l  K_{il}\} - \lambda
K_{ij} - \frac{1}{N}
\end{equation}
and
\begin{equation}
{}\bar{H}_{\{y_i\}} = \frac{1}{2} \sum_i \{a_i,a_i^\dagger\}\qquad.
\end{equation}
It is easy to check that, $[A,a_i] = [A^\dagger,a_i] =
[A^\dagger,a_i^\dagger] = [A^\dagger,a_i] = 0$ and any completely symmetric
state {\it i.e.,} $K_{ij}\psi(\{y_i\}) =\psi(\{y_i\})$ is a solution of both
$\bar{H}_{\{y_i\}}$ and $H_{\{y_i\}}$. the ground-state $|0>$ is annihilated
by $a_i$'s {\it i.e.}, $a_i |0> = 0$.

Keeping in mind the fact that, the spetrum and degeneracies of the CSM match
identically with that of harmonic oscillators, one can choose $(N-1)$
completely symmetrized lowering and raising operators
\begin{equation}
B_n = \sum_i a_i^n\,\,\,\,\,; B_n^\dagger = \sum_i a_i^{\dagger n}\,\,\,
(N\ge n \ge 2)
\end{equation}
such that
\begin{equation}
{}[\bar{H}_{\{y_i\}},B_n(B_n^\dagger)] = - n B_n (n B_n^\dagger)\,\, (n \ge
2)
\qquad.
\end{equation}
For $n = 1$ and $2$, these operators are nothing but the elements of the
harmonic oscillators and the $SU(1,1)$ algebras respectively.$^{12}$ The
$SU(2)$ algebra constructed above for the two 
body example can now be directly generalized to the $N$-particle case. We
construct several non-commuting sets of $SU(2)$ algebras by making use of the
fact that, $A$ and $A^{\dagger}$ commute with all the other $B_n$ and
$B_n^{\dagger}$ operators, for arbitrary $n$. The CCs of $F_n \equiv A^n$ and
$B_n$, obeying the following commutation relations
\begin{equation}
[F_n \,\,,\,\,D_n^\dagger] = [B_n\,\,,\,\, G_{n}^\dagger ] = 1\,\,\,\,,
\end{equation}
can be respectively written in the vacuum sector as
\begin{eqnarray}
D_n^\dagger &=& {1\over n} F_n^\dagger \frac {1}{F_n F_n^\dagger}(A^\dagger A
+ n) \,\,\,,\nonumber\\ G_{n}^\dagger &=& {1\over n} B_n^\dagger \frac
{1}{B_n B_n^\dagger}(\sum_i a_i^\dagger a_i + n)\qquad.
\end{eqnarray}
Using the above operators, the $SU(2)$ generators can be given as
\begin{eqnarray}
{}J_n^+ &=& G_{n}^\dagger F_n \,\,\,,\nonumber\\ {}J_{n}^- &=&
D_{n}^\dagger B_n \,\,\,,\nonumber\\
\mbox{and}\qquad
{}J_n^0 &=&\frac{1}{2} (G_{n}^\dagger B_n - D_{n}^\dagger F_n)
\end{eqnarray}
such that
\begin{eqnarray}
{}[J_n^0 , J_n^\pm] &=& \pm J_n^\pm \,\,\,,\nonumber\\ {}[J_n^+ , J_n^-] &=&
2 J_n^0\qquad.
\end{eqnarray}

Notice that, although the above generators commute with the Hamiltonian, they
are mutually non-commuting due to the fact that $[B_m\,,\,B_n^\dagger] \ne 0$
for $m \ne n$. Hence, the raising and lowering operators of a
given $SU(2)$ only connects the eigenstates of CSM in a given sub-sector of
DE.

Since, we encountered the $B_n$ and $B_n^\dagger$ operators and the fact that
$[B_m\,,\,B_n^\dagger] \ne 0$, it encourages us to study the algebraic
structure underlying these operators. It is obvious that $${}[\bar{H},
[B_n\,,\,B_n^\dagger]] = [\bar{H}, W_{n,n}] = 0$$ and one can check that the
operators $L_n \equiv W_{2,n+1}$ and $L_m \equiv W_{m+1,2}$ generate a
centerless Virasoro algebra. Noticing that ${}[a_i, a_j^\dagger]  =
(\delta_{ij} - \frac{1}{N}) + \lambda \delta_{ij} \sum_l K_{il} - \lambda
K_{ij} $ in relative coordinates $y_i$, we first work out the $W_{m,n}$
commutation relations in $x_i$ coordinates,
the coresponding relations with respect to $y_i$'s can be established by
replacing $\lambda$ independent $\delta_{ij}$ by $\delta_{ij} -
\frac{1}{N}$. One easily checks that,
\begin{eqnarray}
{}[B_m , B_n^\dagger] & = & m \sum_i^N \sum_{r=0}^{n-1} a_i^{\dagger r}
a_i^{m-1}{a_i^{\dagger}}^ {n-1-r} \,\,\,,\nonumber\\ & = & n \sum_i^N
\sum_{s=0}^{m-1} a_i^s a_i^{\dagger {n-1}} a_i^ {m-1-s}
\,\,\,,\nonumber\\
 & \equiv & W_{mn}  \\
\mbox{and}\qquad{}[\bar{H} , W_{mn}] &=& (n - m) W_{mn}\qquad.
\end{eqnarray}
These $W_{mn}$ operators give a new non-linear basis for the $W_\infty$
algebra. First of all,
\begin{equation}
{}[W_{mn} , a_i (a_i^\dagger)] = - m W_{m,n-1}^{(i)} (n W_{m-1,n}^{(i)}) +
M_i
\end{equation}
where $M_i = \sum_j^N \sum_{k=0}^{m-1} \sum_{r=0}^{n-2} a_j^k a_j^{\dagger r}
C_{ij} a_j^{\dagger {n-2-r}} a_j^{m-1-k}$ and $C_{ij} = \lambda \{\delta_{ij}
\sum_l^N K_{il} - K_{ij}\}$.  Using the above equation, $W_{mn}$'s
commutation relations follows as given below:
\begin{eqnarray}
{}[W_{{s-n-1} , {s+n-1}} , W_{{r-m-1} , {r+m-1}}] & = & 2 \{n (r-1) - m
(s-1)\} W_{{s+r-2} , {n+m}} + \cdots \nonumber\\
\end{eqnarray}
where $\cdots$ represent the coupling $\lambda$, $\hbar$ and $M_i$ dependent
non-linear terms. This is a new realization of the $W_\infty$ algebra which
differs from the earlier known ones.$^{13}$

In conclusion, we reanalyzed the two particle CSM and constructed the $SU(2)$
algebra responsible for the degeneracy. This was generalized to the
$N$-particle case and the existence of $N-1$ sets of non-commuting $SU(2)$
algebras was pointed out. In light of this, it is natural to suspect the
existence of a higher symmetry algebra which contains all these $SU(2)$'s as
sub-algebras. In this respect, we arrived at a new $W_\infty$ algebra which
differed from the earlier known ones.$^{13}$ Since, this is a non-linear
algebra, it is difficult to work out its representation theory.
This strongly motivates us to search for the existence of an original
dynamical algebra for CSM. One way out is to look for a linear 
$W_\infty$ algebra; work is in progress in this direction and will be
published elsewhere.

\noindent {\bf Acknowledgements}

N.G would like to thank Dr. J.S. Prakash and E. Harikumar for useful
discussions and U.G.C (India) for the financial support through S.R.F scheme.

\end{document}